\documentclass[11pt,titlepage,a4paper]{article}
\usepackage{bozhomac}  
\usepackage{bozhlogo}  
\usepackage{cite}	

\title{\bfseries	\vspace*{-1.678902345in}
{\huge On angular momentum operator\\[1ex]
       in quantum field theory}
}

\vspace{1.7ex}

\author{
Bozhidar Z.\ Iliev
\thanks{Laboratory of Mathematical Modeling in Physics,
Institute for Nuclear Research and \mbox{Nuclear} Energy,
Bulgarian Academy of Sciences,
Boul.\ Tzarigradsko chauss\'ee~72, 1784 Sofia, Bulgaria}
\thanks{E-mail address: bozho@inrne.bas.bg}
\thanks{URL: http://theo.inrne.bas.bg/$^\sim$bozho/}
}

\date{	
 \vspace{2.27ex}\ShortTitle{Angular momentum operator in QFT}\\[0.27ex]
 \vspace{3.27ex}
\small
	\begin{tabular}{r@{$\colon\to~$}l}
 \vspace{0.09ex} Basic ideas	& July--October, 2001	\\[0.09ex]
 \vspace{0.09ex} Began		& November 1, 2001	\\[0.09ex]
 \vspace{0.09ex} Ended		& November 15, 2001 	\\[0.09ex]
 \vspace{0.09ex} Initial typeset& November 28, 2001--December 1, 2002
							\\[0.09ex]
\vspace{0.09ex} Last update	& November 17, 2002	\\[0.09ex]
 \vspace{0.27ex} Produced	& \fbox{\today}	\\[0.27ex]
	\end{tabular} \\[1.27ex]
\normalsize
	\begin{tabular}{r@{$\colon~$}l}
\vspace{0.27ex} LANL arXiv server E-print No. & hep-th/0211153
	\end{tabular} \\[-0.27ex]
 \vspace{4.27ex}{\Huge\BOZHO}	\\[4.27ex]
 \vspace{0.27ex}\Subject{Quantum field theory}
								\\[2.27ex]
	\begin{tabular}{r@{\hspace{0.512em}}|@{\hspace{0.512em}}l}
 \vspace{0.27ex}\MSC[2000]{81Q99,81T99\\\hspace{0pt}}
&
 \vspace{0.27ex}
	\end{tabular} \\[1.27ex]
 \vspace{0.27ex}\\[0.27ex]
}

\newcommand{\Hil}{\mathcal{F}}		
	\newcommand{\base}{\mathit{M}}	

\newcommand{\ope}[2][{}]{\lindex[\mathcal{#2}]{}{#1}} 
\begin{document}		

\renewcommand{\thefootnote}{\fnsymbol{footnote}} 
\maketitle				
\renewcommand{\thefootnote}{\arabic{footnote}}   

\tableofcontents		


	\begin{abstract}
Relations between two definitions of (total) angular momentum operator, as a
generator of rotations and in the Lagrangian formalism, are explored in
quantum field theory. Generally, these definitions result in different
angular momentum operators, which are suitable for different purposes in the
theory. From the spin and orbital angular momentum  operators (in the
Lagrangian formalism) are extracted additive terms which are conserved
operators and whose sum is the total angular momentum operator.
	\end{abstract}

\section {Introduction}
\label{Introduction}

	Two different definitions of total angular momentum operator in
quantum field theory are in current usage. The one defines it as a conserved
operator arising via the Noether's theorem for rotation\ndash invariant
Lagrangians; we call the arising operator the canonical (or physical) angular
momentum operator, or simply the angular momentum operator. The other one
defines the angular momentum operator as a generator of the representation of
rotations in the Minkowski spacetime on the space of operators acting on the
Hilbert space of some system of quantum fields; we call the so\ndash arising
operator the rotational (or mathematical) angular momentum operator. As we
shall see, this second operator is defined up to a constant second\ndash rank
tensor, which allows its identification with the physical angular momentum
operator on some subset of the Hilbert space of states of a quantum system;
as a rule, that subset is a proper subset.

	The present paper is similar to~\cite{bp-QFT-momentum-operator} and
can be regarded as its continuation.

	The lay-out of the work is as follows.

	In Sect.~\ref{Sect2} is reviewed the notion of angular momentum
operator in the Lagrangian formalism. In Sect.~\ref{Sect3} is considered the
problem of conservation of spin and orbital angular momentum operators. From
these operators are extracted additive parts, which are conserved operators
and whose sum is the (total) angular momentum operator. Sect.~\ref{Sect4}
contains a brief review of the angular momentum operator as a generator of
rotations. In Sect.~\ref{Sect5} are discussed different commutation relations
involving the canonical or rotational angular momentum operators. In
Sect.~\ref{Sect6} is shown that on some set the canonical and rotational
angular momentum operators can coincide, but, generally, these are different
operators. The basic results of the work are summarized in
Sect.~\ref{Conclusion}.

	In what follows, we suppose that there is given a system of quantum
fields, described via field operators $\varphi_i(x)$,
$i=1,\dots,n\in\field[N]$, $x\in\base$ over the 4\ndash dimensional Minkowski
spacetime $\base$ endowed with standard Lorentzian metric tensor
$\eta_{\mu\nu}$ with signature $(+\,-\,-\,-)$.%
\footnote{~%
The quantum fields should be regarded as operator-valued distributions
(acting on a relevant space of test functions) in the rigorous mathematical
setting of Lagrangian quantum field theory. This approach will be considered
elsewhere.%
}
The system's Hilbert space of
states is denoted by $\Hil$ and all considerations are in Heisenberg picture
of motion if the opposite is not stated explicitly. The Greek indices
$\mu,\nu,\dots$ run from 0 to $3=\dim\base-1$ and the Einstein's summation
convention is assumed over indices repeating on different levels. The
coordinates of a point $x\in\base$ are denoted by $x^\mu$, $\bs
x:=(x^1,x^2,x^3)$ and the derivative with respect to $x^\mu$ is
$\frac{\pd}{\pd x^\mu}=:\pd_\mu$. The imaginary unit is denoted by $\iu$ and
$\hbar$ and $c$ stand for the Planck's constant (divided by $2\pi$) and the
velocity of light in vacuum, respectively.


\section{The canonical angular momentum}
	\label{Sect2}

	Most of the material in this section is standard and can be found,
for instance,
in~\cite{Bjorken&Drell-2,Bogolyubov&Shirkov,Roman-QFT,Itzykson&Zuber}.

	Suppose, a system of quantum fields, represented by field operators
$\varphi_i(x)\colon\Hil\to\Hil$, $i=1,\dots,N$, is described by a Lagrangian
$\ope{L}=\ope{L}(x)=\ope{L}(\varphi_i(x),\pd_\mu\varphi_i(x))$ depending on
the fields and their first partial derivatives. Let us introduce the
quantities
	\begin{equation}	\label{2.1}
\pi^{i\mu} :=
\pi^{i\mu}(x) :=
\frac{\pd \ope{L}}{\pd(\pd_\mu \varphi_i(x))},
	\end{equation}
called sometimes generalized momenta. As pointed
in~\cite{bp-QFT-action-principle}, here the derivatives with respect to
$\pd_\mu\varphi_i(x)$, as well as with respect to other generally non\ndash
commuting arguments, should be considered as mappings from some subspace
$\omega$ of the operator space $\{\Hil\to\Hil\}$ over $\Hil$ on
$\{\Hil\to\Hil\}$, \ie
$\pi^{i\mu}\colon\omega\to \{\Hil\to\Hil\}$.%
\footnote{~%
However, in some simple cases, the derivatives with respect to non\ndash
commuting variables may be computed by following the rules of the analysis of
commuting variables by preserving the order of all operators; if this is
the case, one can simply write, e.g., $\pi^{i\mu}\circ\varphi_j$ instead of
$\pi^{i\mu}(\varphi_j)$.%
}

	The system's (canonical) energy\ndash momentum (tensorial) operator is
	\begin{gather}	\label{2.2}
\ope{T}^{\mu\nu}
:= \ope{T}^{\mu\nu}(x)
:=
\sum_{i} \pi^{i\mu}(x) \bigl( \pd^\nu\varphi_i(x) \bigr)
	- \eta^{\mu\nu} \ope{L}(x)
\\\intertext{and satisfies the continuity equation}
				\label{2.3}
\pd_\mu \ope{T}^{\mu\nu} = 0,
\\\intertext{as a result of which the (canonical, dynamical, Noetherian)
momentum operator}
				\label{2.4}
\ope{P}_\mu
:=
\frac{1}{c} \int_{x^0=\const} \ope{T}_{0\mu}(x) \Id^3\bs x
	\end{gather}
is a conserved operator, \ie  $\frac{\od \ope{P}_\mu}{\od x^0}=0$.

	Suppose under a 4-rotation
 $x^\mu\mapsto x^{\prime \mu} = x^\mu + \varepsilon^{\mu\nu}x_\nu$, with
$x_\nu:=\eta_{\nu\mu} x^\mu$ and antisymmetric real parameters
$\varepsilon^{\mu\nu}=-\varepsilon^{\nu\mu}$, the field operators transform
as
\(
\varphi_i(x)  \mapsto \varphi'_i(x')
:=
\varphi_i(x) +
\sum_{\mu<\nu} I_{i\mu\nu}^{j} \varphi_j(x) \varepsilon^{\mu\nu}
+ \dotsb,
\)
where the dots stand for second and higher order terms in
$\varepsilon^{\mu\nu}$ and the numbers $I_{i\mu\nu}^{j}=-I_{i\nu\mu}^{j}$
characterize the behaviour of the field operators under rotations.

	The total angular momentum density operator of a system of quantum
fields $\varphi_i(x)$ is
	\begin{align}	\label{2.5}
& \ope{M}_{\mu\nu}^{\lambda}(x)
= \ope{L}_{\mu\nu}^{\lambda}(x) + \ope{S}_{\mu\nu}^{\lambda}(x)
\\\intertext{where}
			\label{2.6}
& \ope{L}_{\mu\nu}^{\lambda}(x)
:=
x_\mu \Sprindex[\ope{T}]{\nu}{\lambda}(x)
-
x_\nu \Sprindex[\ope{T}]{\mu}{\lambda}(x)
\\			\label{2.7}
& \ope{S}_{\mu\nu}^{\lambda}(x)
:=
\sum_{i,j} \pi^{i\lambda}(x)\bigl( \varphi_j(x) \bigr) I_{i\mu\nu}^{j}
	\end{align}
are respectively the orbital and spin angular momentum density operators. As
a result of the continuity equation
	\begin{equation}	\label{2.8}
\pd_\lambda \ope{M}_{\mu\nu}^{\lambda}(x) = 0,
	\end{equation}
the (total) angular momentum operator
	\begin{align}	\label{2.9}
& \ope{M}_{\mu\nu} = \ope{L}_{\mu\nu}(x^0) + \ope{S}_{\mu\nu}(x^0)
\\\intertext{where}
			\label{2.10}
& \ope{L}_{\mu\nu}(x^0)
:=
\frac{1}{c} \int_{x^0=\const} \ope{L}_{\mu\nu}^0(x) \Id^3\bs x
\\			\label{2.11}
& \ope{S}_{\mu\nu}(x^0)
:=
\frac{1}{c} \int_{x^0=\const} \ope{S}_{\mu\nu}^0(x) \Id^3\bs x
	\end{align}
are respectively the orbital and spin angular momentum operators, is a
conserved quantity, \ie
	\begin{equation}	\label{2.12}
\frac{\od \ope{M}_{\mu\nu}}{\od x^0} = 0 .
	\end{equation}
Notice, in the general case, the operators~\eref{2.10} and~\eref{2.11} are
not conserved (see below Sect.~\ref{Sect3}).


\section {Conservation laws}
\label{Sect3}

	Since from~\eref{2.3} and~\eref{2.5}--\eref{2.8} follow the equations
	\begin{align}
			\label{3.1}
& \pd_\lambda \ope{L}_{\mu\nu}^{\lambda} = \ope{T}_{\mu\nu} - \ope{T}_{\nu\mu}
\\			\label{3.2}
& \pd_\lambda \ope{S}_{\mu\nu}^{\lambda}
	+ \pd_\lambda \ope{L}_{\mu\nu}^{\lambda} = 0 ,
	\end{align}
in the general case, when the (canonical) energy-momentum tensor
$\ope{T}_{\mu\nu}$ is non\ndash symmetric,%
\footnote{~%
By adding to~\eref{2.2} a full divergence term, one can form a symmetric
energy\ndash momentum tensor; for details,
see~\cite[sec.~2 and the references therein]{Pauli-RelativisticTheories}.
However, this does not change anything in our conclusions as expressions,
like the r.h.s.\ of~\eref{3.1} with $\ope{T}_{\mu\nu}$ defined by~\eref{2.2},
remain the same if one works with the new symmetric tensor.%
}
the spin and orbital angular momentum operators are not conserved. However,
from the operators~\eref{2.10} and~\eref{2.11} can be extracted additive
conserved ones, which are, in fact, the invariants characteristics of the
spin and orbital angular properties of quantum systems.

	For the purpose, define the antisymmetric operators
	\begin{equation}	\label{3.3}
t_{\mu\nu}(x)
:=
\int\limits_{x_0^0}^{x^0}
\bigl( \ope{T}_{\mu\nu}(x) - \ope{T}_{\nu\mu}(x) \bigr) \Id x^0
=
- t_{\nu\mu}
	\end{equation}
with $x_0^0$ being some arbitrarily fixed instant of the time coordinate
$x^0$. Let us put
	\begin{subequations}	\label{3.4}
	\begin{align}	\label{3.4a}
\lindex[\mspace{-6mu}{\ope{S}_{\mu\nu}}]{}{0}
:& =
\frac{1}{c} \int_{x^0=\const}
	\{ \ope{S}_{\mu\nu}^{0}(x) + t_{\mu\nu}(x) \} \Id^3\bs x
 =
\ope{S}_{\mu\nu}(x_0) + \frac{1}{c} \int_{x^0=\const}
	t_{\mu\nu}(x) \Id^3\bs x
\\			\label{3.4b}
\lindex[\mspace{-6mu}{\ope{L}_{\mu\nu}}]{}{0}
: & =
\frac{1}{c} \int_{x^0=\const}
	\{ \ope{L}_{\mu\nu}^{0}(x) - t_{\mu\nu}(x) \} \Id^3\bs x
 =
\ope{L}_{\mu\nu}(x_0) - \frac{1}{c} \int_{x^0=\const}
	t_{\mu\nu}(x)  \Id^3\bs x .
	\end{align}
	\end{subequations}
Since
	\begin{gather}	\label{3.4-1}
t_{\mu\nu}(x) = 0
\qquad\text{for } \ope{T}_{\mu\nu}=\ope{T}_{\nu\mu},
\\\intertext{we have}
			\label{3.4-2}
\lindex[\mspace{-6mu}{\ope{S}_{\mu\nu}}]{}{0} = \ope{S}_{\mu\nu}(x_0)
\quad
\lindex[\mspace{-6mu}{\ope{L}_{\mu\nu}}]{}{0} = \ope{L}_{\mu\nu}(x_0)
\qquad\text{for } \ope{T}_{\mu\nu}=\ope{T}_{\nu\mu}.
	\end{gather}
Applying~\eref{3.1}--\eref{3.3}, we get
	\begin{equation*}
\frac{\od}{\od x^0} \lindex[\mspace{-6mu}{\ope{L}_{\mu\nu}}]{}{0}
=
\frac{1}{c} \int_{x^0=\const} \Id^3 \bs x
\{ \pd_0 \ope{L}_{\mu\nu}^{0}(x) - \pd_0 t_{\mu\nu}(x) \}
 =
- \frac{1}{c} \int_{x^0=\const} \Id^3 \bs x
\sum_{a=1}^{3} \pd_a \ope{L}_{\mu\nu}^{a}(x)
	\end{equation*}
\vspace{-4ex}
	\begin{multline*}
\frac{\od}{\od x^0} \lindex[\mspace{-6mu}{\ope{S}_{\mu\nu}}]{}{0}
=
\frac{1}{c} \int_{x^0=\const} \Id^3 \bs x
\{ \pd_0 \ope{S}_{\mu\nu}^{0}(x) + \pd_0 t_{\mu\nu}(x) \}
\\ =
\frac{1}{c} \int_{x^0=\const} \Id^3 \bs x
\Bigl\{
- \sum_{a=1}^{3} \pd_a \ope{S}_{\mu\nu}^{a}(x)
- \pd_\lambda \ope{L}_{\mu\nu}^{\lambda}(x)
+ \ope{T}_{\mu\nu}(x) - \ope{T}_{\nu\mu}(x)
\Bigr\}
\\ =
- \frac{1}{c} \int_{x^0=\const} \Id^3 \bs x
\sum_{a=1}^{3} \pd_a \ope{S}_{\mu\nu}^{a}(x) .
	\end{multline*}
Let us suppose that the field operators tend to zero sufficiently fast at
spacial infinity and
$\ope{L}_{\mu\nu}^{a}(x), \ope{S}_{\mu\nu}^{a}(x) \to 0$
when $x$ tends to (some) spacial infinity. Then, from the last equalities, we
derive the conservation laws
	\begin{equation}	\label{3.5}
\frac{\od}{\od x^0} \lindex[\mspace{-6mu}{\ope{L}_{\mu\nu}}]{}{0} = 0
\qquad
\frac{\od}{\od x^0} \lindex[\mspace{-6mu}{\ope{S}_{\mu\nu}}]{}{0} = 0 .
	\end{equation}
Thus, the operators~\eref{3.4} are conserved. Besides, due to
equations~\eref{2.9} and~\eref{3.4}, their sum is exactly the angular
momentum operator,
	\begin{equation}	\label{3.6}
\ope{M}_{\mu\nu}
=
  \lindex[\mspace{-6mu}{\ope{L}_{\mu\nu}}]{}{0}
+ \lindex[\mspace{-6mu}{\ope{S}_{\mu\nu}}]{}{0} .
	\end{equation}
Moreover, if one starts from the
definitions~\eref{3.3},~\eref{3.4}, \eref{2.1}--\eref{2.7},
and~\eref{2.9}--\eref{2.11}, one can prove ~\eref{3.5} via a direct
calculation (involving the field equations) the validity of~\eref{3.5} and,
consequently, the conservation law~\eref{2.12} becomes a corollary of the
ones for $\lindex[\mspace{-6mu}{\ope{L}_{\mu\nu}}]{}{0}$ and
$\lindex[\mspace{-6mu}{\ope{S}_{\mu\nu}}]{}{0}$.

 	Since the operator $\lindex[\mspace{-6mu}{\ope{S}_{\mu\nu}}]{}{0}$
characterizes entirely internal properties of the considered system of
quantum fields, it is suitable to be called its \emph{spin (or spin charge)
operator}. Similar name, the \emph{orbital operator}, is more or less
applicable for $\lindex[\mspace{-6mu}{\ope{L}_{\mu\nu}}]{}{0}$ too. Particular examples of
these quantities will be presented elsewhere.

	The above considerations are, evidently, true in the case of
classical Lagrangian formalism of commuting variables too.%
\footnote{~%
The only essential change in this case is that expressions like
$\pi^{i\mu}(\varphi_j)$ should be replaced with $\pi^{i\mu}\varphi_j$, where
an ordinary multiplication between functions is understood.%
}


\section {The generators of rotations}
\label{Sect4}

	Besides~\eref{2.5}, there is a second definition of the total angular
momentum operator, which defines it as a generator of the representation of
rotational subgroup of Poincar\'e group on the space of operators acting on
the Hilbert space $\Hil$ of the fields $\varphi_i(x)$.%
\footnote{~%
In axiomatic quantum field theory, this is practically the only definition of
(total) angular momentum; see, e.g.,~\cite[sec.~3.1.2]{Itzykson&Zuber},
\cite[p.~146]{Bogolyubov&et_al.-AxQFT}
and~\cite[sec.~7.1]{Bogolyubov&et_al.-QFT}. This definition can also be found
the (text)books on Lagrangian/canonical quantum field theory; see, for
instance,~\cite[sec.~2.1]{Roman-QFT}, \cite[sec.~3.1.2]{Itzykson&Zuber},
\cite[\S~68]{Bjorken&Drell-2}, and~\cite[sec.~9.4]{Bogolyubov&Shirkov}.%
}
The so\ndash arising operator will be referred as the \emph{rotational} (or
mathematical) \emph{angular momentum operator} and will be denoted by
$\ope{M}_{\mu\nu}^{\text{r}}$. It is defined as follows. If $x\mapsto x'$,
with
\(
x^{\prime\lambda}
= x^\lambda + \varepsilon^{\lambda\nu}x_\nu
= x^\lambda + \sum_{\mu<\nu} \varepsilon^{\mu\nu}
	( \delta_\mu^\lambda x_\nu - \delta_\nu^\lambda x_\mu ),
\)
 $\varepsilon^{\mu\nu} = - \varepsilon^{\nu\mu}$,
is a rotation of the Minkowski spacetime, then it induces the transformation
	\begin{equation}	\label{4.1}
\ope{A}(x)\mapsto \ope{A}(x')
=:
\e^{ -\iih\frac{1}{2} \varepsilon^{\mu\nu} \ope{M}_{\mu\nu}^{\text{r}}  }
\circ \ope{A}(x) \circ
\e^{ \iih\frac{1}{2} \varepsilon^{\mu\nu} \ope{M}_{\mu\nu}^{\text{r}} } ,
	\end{equation}
where $\circ$ denotes composition of mappings, $\ope{A}(x)\colon\Hil\to\Hil$
is a linear operator and $\ope{M}_{\mu\nu}^{\text{r}}\colon\Hil\to\Hil$ are
some operators. In differential form, equation~\eref{4.1} is equivalent to
	\begin{equation}	\label{4.2}
[ \ope{A}(x), \ope{M}_{\mu\nu}^{\text{r}} ]_{\_}
=
\ih \Bigl(
x_\mu \frac{\pd \ope{A}(x)}{\pd x^\nu} - x_\nu \frac{\pd \ope{A}(x)}{\pd x^\mu}
\Bigr),
	\end{equation}
where
$[\ope{A},\ope{B}]_{\_}:=\ope{A}\circ \ope{B}-\ope{B}\circ \ope{A}$
is the commutator of $\ope{A},\ope{B}\colon\Hil\to\Hil$.

	However, the behaviour of the field operators $\varphi_i(x)$ under
rotations is, generally, more complicated than~\eref{4.1},
\viz~\cite{Bjorken&Drell-2,Bogolyubov&Shirkov}
	\begin{equation}	\label{4.3}
\varphi_i(x)\mapsto
\sum_{j} \bigl( S^{-1} \bigr) \! \Sbrindex{i}{j} (\varepsilon) \varphi_j(x')
=
\e^{ -\iih\frac{1}{2} \varepsilon^{\mu\nu} \ope{M}_{\mu\nu}^{\text{r}} }
\circ \varphi_i(x) \circ
\e^{ \iih\frac{1}{2} \varepsilon^{\mu\nu} \ope{M}_{\mu\nu}^{\text{r}} },
	\end{equation}
with $S=[\Sbrindex[S]{i}{j}(\varepsilon)]$ being a depending on
$\varepsilon=[\varepsilon^{\mu\nu}]$ non\ndash degenerate matrix. The
appearance of a, generally, non\ndash unit matrix $S$ in~\eref{4.3} is due to
the fact that under the set of field operators $\{\varphi_i(x)\}$ is
understood the collection of all of the \emph{components} $\varphi_i(x)$ of
the fields forming a given system of quantum fields. This means that if, say,
the field operators $\varphi_{i_1}(x),\dots,\varphi_{i_n}(x)$ for some indices
$i_1,\dots,i_n$, $n\in\field[N]$, represent a particular quantum field, they
are components of an operator vector (a vector\ndash valued operator) $\phi$
with respect to some basis $\{f_{i_1},\dots,f_{i_n}\}$ of operator
vector space to which $\phi$ belongs,
\(
\phi(x)
= \varphi_{i_1}(x) f_{i_1} + \dots + \varphi_{i_n}(x) f_{i_n}.
\)
Under rotations, $\phi(x)$ transforms according to~\eref{4.1}, but its
components $\varphi_{i_1}(x),\dots,\varphi_{i_n}(x)$ generally do not; they
transform in conformity with~\eref{4.3} because a change $x\mapsto x'$ is
supposed to induce a linear change
\(
f_{i_\alpha} \mapsto f'_{i_\alpha}
=
\sum_{\beta=1}^{n}
\Sbrindex[{ \bigl(\bigl( S^\phi\bigr)^{-1}\bigr) }]{i_\alpha}{i_\beta}
(\varepsilon)
f_{i_\beta},
\)
 $\alpha=1,\dots,n$,
with a non\ndash degenerate matrix
\(
S^\phi(\varepsilon)
:= [ \Sbrindex[{S^\phi}]{i_\alpha}{i_\beta} (\varepsilon) ]
\)%
\footnote{~%
The matrix $S(\varepsilon)$ in~\eref{4.3} is a direct sum of the matrices
$S^\phi(\varepsilon)$ for the independent fields $\phi$ forming the system
under consideration.%
}
and
\(
\phi(x')
= \varphi_{i_1}(x') f'_{i_1} + \dots + \varphi_{i_n}(x') f'_{i_n}.
\)

	Often~\eref{4.3} is written in a differential form
as~\cite[eq.~(11.73)]{Bjorken&Drell-2}
	\begin{equation}	\label{4.4}
[ \varphi_i(x), \ope{M}_{\mu\nu}^{\text{r}} ]_{\_}
=
\ih \Bigl(
  x_\mu \frac{\pd \varphi_i(x)}{\pd x^\nu}
- x_\nu \frac{\pd \varphi_i(x) }{\pd x^\mu}
\Bigr)
+ \ih \sum_{j} I_{i\mu\nu}^{j} \varphi_j(x) ,
	\end{equation}
where $I_{i\mu\nu}^{j}=-I_{i\nu\mu}^{j}$ are defined by
\(
I_{i\mu\nu}^{j}
:=
\frac{ \pd \Sbrindex[S]{i}{j}(\varepsilon) } {\pd\varepsilon^{\mu\nu}}
\Big|_{\varepsilon=0}
\)
for $\mu<\nu$, \ie
\(
\Sbrindex[S]{i}{j}(\varepsilon)
=
\delta_i^j + \frac{1}{2} I_{i\mu\nu}^{j} \varepsilon^{\mu\nu} + \dotsb
\)
with $\delta_i^j$ being the Kroneker delta\ndash symbol and the dots stay for
higher order terms in $\varepsilon^{\mu\nu}$.

	There is a simple relation between $\ope{M}_{\mu\nu}^{\text{r}}$ and
the rotation operator on $\Hil$. Let $\ope{M}_{\mu\nu}^{\text{QM}}$, where QM
stands for Quantum Mechanics (see below), denotes the Hermitian generator of
rotations in $\Hil$, \ie if $\ope{X}(x)\in\Hil$, then
	\begin{gather}	\label{4.5}
\ope{X}(x) \mapsto
\ope{X}(x')
=
\e^{ \iih\frac{1}{2} \varepsilon^{\mu\nu} \ope{M}_{\mu\nu}^{\text{QM}} }
	(\ope{X}(x))
\\\intertext{with $x^{\prime\mu}=x^\mu+\varepsilon^{\mu\nu}x_\nu$.
Explicitly, we have
(see~\cite{Itzykson&Zuber}, \cite[eq.~(6.5)]{Bogolyubov&et_al.-AxQFT}
or~\cite[eq.~(7.14)]{Bogolyubov&et_al.-QFT}) }
			\label{4.6}
\ope{M}_{\mu\nu}^{\text{QM}}
= \ih ( x_\mu \pd_\nu - x_\nu\pd_\mu ),
	\end{gather}
which  is exactly the orbital angular momentum  operator in quantum
mechanics if one restricts $\mu$ and $\nu$ to the range $1,2,3$, forms the
corresponding to~\eref{4.6} 3\ndash vector operator (see,
e.g.,~\cite{Messiah-1}) and identifies $\Hil$ with the Hilbert space of
states of quantum mechanics. The equalities
	\begin{align}	\label{4.7}
\ope{A}(x')
& =
\e^{ \iih\frac{1}{2} \varepsilon^{\mu\nu} \ope{M}_{\mu\nu}^{\text{QM}}  }
\circ \ope{A}(x) \circ
\e^{-\iih\frac{1}{2} \varepsilon^{\mu\nu} \ope{M}_{\mu\nu}^{\text{QM}} }
\\			\label{4.8}
\sum_{j} \bigl( S^{-1} \bigr) \! \Sbrindex{i}{j} (\varepsilon) \varphi_j(x')
& =
\e^{ \iih\frac{1}{2} \varepsilon^{\mu\nu} \ope{M}_{\mu\nu}^{\text{QM}} }
\circ \varphi_i(x) \circ
\e^{-\iih\frac{1}{2} \varepsilon^{\mu\nu} \ope{M}_{\mu\nu}^{\text{QM}} }
	\end{align}
are simple corollaries of~\eref{4.5} and in differential form read
	\begin{align}	\label{4.9}
[ \ope{A}(x), \ope{M}_{\mu\nu}^{\text{QM}} ]_{\_}
& =
- \ih \Bigl(
  x_\mu \frac{\pd \ope{A}(x)}{\pd x^\nu}
- x_\nu \frac{\pd \ope{A}(x)}{\pd x^\mu}
\Bigr),
\\			\label{4.10}
[ \varphi_i(x), \ope{M}_{\mu\nu}^{\text{QM}} ]_{\_}
& =
- \ih \Bigl(
  x_\mu \frac{\pd \varphi_i(x)}{\pd x^\nu}
- x_\nu \frac{\pd \varphi_i(x) }{\pd x^\mu}
\Bigr)
- \ih \sum_{j} I_{i\mu\nu}^{j} \varphi_j(x) .
	\end{align}
Comparing the last equations with~\eref{4.2} and~\eref{4.4}, we find
	\begin{subequations}	\label{4.11}
	\begin{align}	\label{4.11a}
[ \ope{A}(x) , \ope{M}_{\mu\nu}^{\text{QM}}
		+ \ope{M}_{\mu\nu}^{\text{r}} ]_{\_} & = 0
\\			\label{4.11b}
[ \varphi_i(x) , \ope{M}_{\mu\nu}^{\text{QM}}
		+ \ope{M}_{\mu\nu}^{\text{r}} ]_{\_} & = 0 .
	\end{align}
	\end{subequations}
If we admit~\eref{4.11a} (or~\eref{4.2}) to hold for any
$\ope{A}(x)\colon\Hil\to\Hil$, the Schur's lemma%
\footnote{~%
See, e.g,~\cite[appendix~II]{Rumer&Fet}, \cite[sec.~8.2]{Kirillov-1976},
\cite[ch.~5, sec.~3]{Barut&Roczka}.%
}
implies
	\begin{equation}	\label{4.12}
\ope{M}_{\mu\nu}^{\text{r}}
= - \ope{M}_{\mu\nu}^{\text{QM}} + m_{\mu\nu} \id_\Hil
= - \ih(x_\mu\pd_\nu-x_\nu\pd_\mu) + m_{\mu\nu} \id_\Hil ,
	\end{equation}
where $\id_\Hil$ is the identity mapping of $\Hil$ and $m_{\mu\nu}$ are real
numbers with dimension of angular momentum and forming the covariant
components of some tensor of second rank.

	One should be aware of the fact that the notation
$\ope{M}_{\mu\nu}^{\mathrm{QM}}$ for the generator of rotations on $\Hil$
only emphasizes on the analogy with a similar operator in quantum mechanics
(see also equation~\eref{4.6}); however, these two operators are
completely different as they act on different spaces, the Hilbert space of
states of quantum field theory and quantum mechanics respectively, which
cannot be identified. For that reason, we cannot say that~\eref{4.12} with
$m_{\mu\nu}=0$ implies that if the angular momentum of a system in quantum
field theory and in quantum mechanics are equal up to a sign.


\section {Discussion}
\label{Sect5}

	The problem for coincidence of the both definitions of (total)
angular momentum operator, the canonical and as generator of rotations, is a
natural one and its positive answer is, more or less, implicitly assumed in
the literature~\cite{Roman-QFT,Itzykson&Zuber}. However, these definitions
originate from different approaches to quantum field theory: the canonical
is due to the Lagrangian
formalism~\cite{Bogolyubov&Shirkov,Bjorken&Drell,Itzykson&Zuber}, while the
another one finds its natural place in axiomatic quantum field
theory~\cite{Bogolyubov&et_al.-AxQFT,Bogolyubov&et_al.-QFT}.

	As a condition weaker than
	\begin{equation}	\label{5.1}
\ope{M}_{\mu\nu} = \ope{M}_{\mu\nu}^{\text{r}} ,
	\end{equation}
the relation~\eref{4.4}, or its integral version~\eref{4.3}, is assumed with
$\ope{M}_{\mu\nu}$ for $\ope{M}_{\mu\nu}^{\text{r}}$, \ie
	\begin{equation}	\label{5.2}
[ \varphi_i(x), \ope{M}_{\mu\nu} ]_{\_}
=
\ih \Bigl(
  x_\mu \frac{\pd \varphi_i(x)}{\pd x^\nu}
- x_\nu \frac{\pd \varphi_i(x) }{\pd x^\mu}
\Bigr)
+ \ih \sum_{j} I_{i\mu\nu}^{j} \varphi_j(x) .
	\end{equation}
For instance, in~\cite[\S~68]{Bjorken&Drell-2} or
in~\cite[sections~9.3 and~9.4]{Bogolyubov&Shirkov}, the last equation is
proved under some explicitly written conditions concerning the transformation
properties of the field operators and state vectors under Poincar\'e
transformations. But, whatever the conditions leading to~\eref{5.2} are, they
and~\eref{5.2} are external to the Lagrangian formalism by means of which the
canonical angular momentum operator $\ope{M}_{\mu\nu}$ is defined. Usually,
equation~\eref{5.2} is considered as one of the conditions ensuring the
relativistic covariance of the Lagrangian formalism and the theory is
restricted to Lagrangians for which~\eref{5.2} is a consequence of the field
equations (and, possibly, some restrictions on the formalism).%
\footnote{~%
See~\cite{Bjorken&Drell-2}, especially the comments on this item in
section~68 of this book.%
}

	It should be noted, the more general equation~\eref{4.2} with
$\ope{M}_{\mu\nu}$ for $\ope{M}_{\mu\nu}^{\text{r}}$ and arbitrary operator
$\ope{A}(x)$, \ie
	\begin{equation}	\label{5.3}
[ \ope{A}(x), \ope{M}_{\mu\nu} ]_{\_}
=
\ih \Bigl(
  x_\mu \frac{\pd \ope{A}(x)}{\pd x^\nu}
- x_\nu \frac{\pd \ope{A}(x)}{\pd x^\mu}
\Bigr),
	\end{equation}
cannot be valid; a simple counter example is provided by
$\ope{A}(x)=\ope{P}_\mu$, with $\ope{P}_\mu$ being the canonical momentum
operator of the considered system of quantum fields, for which the r.h.s.\
of~\eref{5.3} vanishes, as $\pd_\lambda \ope{P}_\mu=0$, but
\(
[ \ope{M}_{\mu\nu}, \ope{P}_\lambda ]_{\_}
=
\ih ( \eta_{\lambda\mu} \ope{P}_\nu - \eta_{\lambda\nu} \ope{P}_\mu )
\)
(see, e.g.,~\cite[eq.~(2-83)]{Roman-QFT}
or~\cite[eq.~6.1]{Bogolyubov&et_al.-AxQFT}). However, if it
happens~\eref{5.3} to hold for operators $\ope{A}(x)$ that form an irreducible
unitary representation of some group, then, combining~\eref{5.3}
and~\eref{4.2} and applying the Schur lemma, we get
	\begin{equation}	\label{5.4}
\ope{M}_{\mu\nu}
=   \ope{M}_{\mu\nu}^{\text{r}}  + n_{\mu\nu} \id_\Hil
= - \ope{M}_{\mu\nu}^{\text{QM}} + (m_{\mu\nu} + n_{\mu\nu}) \id_\Hil ,
	\end{equation}
where~\eref{4.12} was used and $n_{\mu\nu}$ are constant covariant components
of some second\ndash rank tensor.

	Defining the operator
	\begin{equation}	\label{5.5}
\ope{P}_{\mu}^{\text{QM}} := \ih \frac{\pd}{\pd x^\mu},
	\end{equation}
which is the 4-dimensional analogue of the momentum operator in quantum
mechanics, we see that it and $\ope{M}_{\mu\nu}^{\text{QM}}$ (see~\eref{4.6})
satisfy the next relations
	\begin{subequations}	\label{5.6}
	\begin{align}
			\label{5.6a}
[ \ope{P}_{\mu}^{\text{QM}} , \ope{P}_{\nu}^{\text{QM}} ]_{\_}
& = 0
\\			\label{5.6b}
[ \ope{M}_{\mu\nu}^{\text{QM}} , \ope{P}_{\lambda}^{\text{QM}} ]_{\_}
& = -\ih ( \eta_{\lambda\mu} \ope{P}_{\nu}^{\text{QM}}
	 - \eta_{\lambda\nu} \ope{P}_{\mu}^{\text{QM}} )
\\			\label{5.6c}
[ \ope{M}_{\mu\nu}^{\text{QM}} , \ope{M}_{\kappa\lambda}^{\text{QM}} ]_{\_}
& = -\ih
\bigl(
  \eta_{\mu\kappa} \ope{M}_{\nu\lambda}^{\text{QM}}
+ \eta_{\kappa\nu} \ope{M}_{\lambda\mu}^{\text{QM}}
+ \eta_{\nu\lambda} \ope{M}_{\mu\kappa}^{\text{QM}}
+ \eta_{\lambda\mu} \ope{M}_{\kappa\nu}^{\text{QM}}
\bigr) ,
	\end{align}
	\end{subequations}
which characterize the Lie algebra of the Poincar\'e
group (see~\cite[sec.~6.1]{Bogolyubov&et_al.-AxQFT},
\cite[sec.~7.1]{Bogolyubov&et_al.-QFT}, and~\cite{Ramond-FT}). From~\eref{5.6}
it is easily seen, the operators~\eref{4.12} and
	\begin{equation}	\label{5.7}
\ope{P}_{\mu}^{\text{t}} = - \ope{P}_{\mu}^{\text{QM}}  + p_\mu \id_\Hil ,
	\end{equation}
where $p_\mu$ are constant covariant components of a 4\ndash vector, satisfy
the Lie algebra of the Poincar\'e group
(see~\cite[sec.~6.1]{Bogolyubov&et_al.-AxQFT},
\cite[sec.~7.1]{Bogolyubov&et_al.-QFT} and~\cite[pp.~76--78]{Roman-QFT}), \ie
	\begin{subequations}	\label{5.8}
	\begin{align}
			\label{5.8a}
[ \ope{P}_{\mu}^{\text{t}} , \ope{P}_{\nu}^{\text{t}} ]_{\_}
& = 0
\\			\label{5.8b}
[ \ope{M}_{\mu\nu}^{\text{r}} , \ope{P}_{\lambda}^{\text{t}} ]_{\_}
& = \ih ( \eta_{\lambda\mu} \ope{P}_{\nu}^{\text{t}}
	 - \eta_{\lambda\nu} \ope{P}_{\mu}^{\text{t}} )
\\			\label{5.8c}
[ \ope{M}_{\mu\nu}^{\text{r}} , \ope{M}_{\kappa\lambda}^{\text{r}} ]_{\_}
& = \ih
\bigl(
  \eta_{\mu\kappa} \ope{M}_{\nu\lambda}^{\text{r}}
+ \eta_{\kappa\nu} \ope{M}_{\lambda\mu}^{\text{r}}
+ \eta_{\nu\lambda} \ope{M}_{\mu\kappa}^{\text{r}}
+ \eta_{\lambda\mu} \ope{M}_{\kappa\nu}^{\text{r}}
\bigr)
	\end{align}
	\end{subequations}
if and only if
	\begin{equation}	\label{5.9}
p_\mu = 0 \quad m_{\mu\nu} = 0 .
	\end{equation}
Thus, the relations~\eref{5.8} remove completely the arbitrariness in the
operators $\ope{P}_{\mu}^{\text{t}}$ and $\ope{M}_{\mu\nu}^{\text{r}}$.
The equations~\eref{5.8} are often~\cite{Roman-QFT} assumed to hold for the
canonical momentum and angular momentum operators,

	\begin{subequations}	\label{5.10}
	\begin{align}
			\label{5.10a}
[ \ope{P}_{\mu} , \ope{P}_{\nu} ]_{\_}
& = 0
\\			\label{5.10b}
[ \ope{M}_{\mu\nu} , \ope{P}_{\lambda} ]_{\_}
& = \ih ( \eta_{\lambda\mu} \ope{P}_{\nu}
	 - \eta_{\lambda\nu} \ope{P}_{\mu}
\\			\label{5.10c}
[ \ope{M}_{\mu\nu} , \ope{M}_{\kappa\lambda} ]_{\_}
& = \ih
\bigl(
  \eta_{\mu\kappa} \ope{M}_{\nu\lambda}
+ \eta_{\kappa\nu} \ope{M}_{\lambda\mu}
+ \eta_{\nu\lambda} \ope{M}_{\mu\kappa}
+ \eta_{\lambda\mu} \ope{M}_{\kappa\nu}
\bigr) .
	\end{align}
	\end{subequations}
However, these equations, as well as~\eref{5.2}, are external to the
Lagrangian formalism and, consequently, their validity should be checked for
any particular Lagrangian.%
\footnote{~%
Elsewhere we shall demonstrate that equation~\eref{5.10b} is not valid for a
free spin~0, $\frac{1}{2}$ and~1 fields; more precisely, it holds, for these
fields, with an opposite sign of its r.h.s., \ie with $-\ih$ instead of
$\ih$.%
}


\section {Inferences}
\label{Sect6}

	Regardless of the fact that the equation~\eref{5.2} holds in most
cases, it is quite different from the similar to it relation~\eref{4.4}.
Indeed, the relation~\eref{4.4} is an identity with respect to the field
operators $\varphi_i(x)$, while~\eref{5.2} can be considered as an equation
with respect to them. Thus,~\eref{5.2} can be considered as equations of
motion relative to the field operators (known as (part of) the Heisenberg
equations/relations), but~\eref{4.4} are identically valid with respect to
these operators. Consequently, from this position, the possible
equality~\eref{5.1} is unacceptable because it will entail~\eref{5.2} as an
identity regardless of the Lagrangian one starts off.

	Let $\ope{X}\in\Hil$ be a state vector of the considered system of
quantum fields. Since we work in Heisenberg picture, it is a constant vector
and, consequently, we have
	\begin{gather}	\label{6.1}
\ope{M}_{\mu\nu}^{\text{QM}}(\ope{X}) = 0
\quad
\e^{ \iih\frac{1}{2} \varepsilon^{\mu\nu} \ope{M}_{\mu\nu}^{\text{QM}} } (x)
= 0
\quad
\ope{P}_{\mu}^{\text{QM}}(\ope{X}) = 0 .
\\\intertext{Then~\eref{4.12} implies}
			\label{6.2}
\ope{M}_{\mu\nu}^{\text{r}}(\ope{X}) = m_{\mu\nu} \ope{X} .
	\end{gather}
As we intend to interpret $\ope{M}_{\mu\nu}^{\text{r}}$ as system's total
angular momentum operator, the last equation entails the \emph{interpretation
of $m_{\mu\nu}$ as components of the total 4\ndash angular momentum of the
system}. To justify this interpretation, one should assume $\ope{X}$ to be
also an eigenvector of the canonical angular momentum operator with the same
eigenvalues, \ie
	\begin{equation}	\label{6.3}
\ope{M}_{\mu\nu}(\ope{X}) = m_{\mu\nu} \ope{X} .
	\end{equation}
From here two conclusions can be made. On one hand, the equality~\eref{5.4}
may be valid only for $n_{\mu\nu}=0$, but, as we said earlier, this equation
cannot hold in the general case. On other hand, equations~\eref{6.2}
and~\eref{6.3} imply
	\begin{equation}	\label{6.4}
  \ope{M}_{\mu\nu} \big|_{\ope{D}_m}
= \ope{M}_{\mu\nu}^{\text{r}} \big|_{\ope{D}_m}
	\end{equation}
where
	\begin{equation}	\label{6.5}
\ope{D}_m
:= \{ \ope{X}\in\Hil : \ope{M}_{\mu\nu}(\ope{X}) = m_{\mu\nu} \ope{X} \} .
	\end{equation}
Generally, the relation~\eref{6.4} is weaker than~\eref{5.1} and implies it
if a basis of $\Hil$ can be formed from vectors in $\ope{D}_m$.

	An alternative to~\eref{6.4} and~\eref{5.1} are the equalities
	\begin{equation}	\label{6.6}
[ \varphi_i(x), \ope{M}_{\mu\nu} ]_{\_}
=
[ \varphi_i(x), \ope{M}_{\mu\nu}^{\text{r}} ]_{\_}
 =
\ih \Bigl(
  x_\mu \frac{\pd \varphi_i(x)}{\pd x^\nu}
- x_\nu \frac{\pd \varphi_i(x) }{\pd x^\mu}
\Bigr)
+ \ih \sum_{j} I_{i\mu\nu}^{j} \varphi_j(x) ,
	\end{equation}
which do not imply~\eref{5.1}. The above discussion also shows that the
equality $\ope{M}_{\mu\nu}^{\text{r}}=-\ope{M}_{\mu\nu}^{\text{QM}}$ can be
valid only for states with vanishing total angular momentum.

	In Sect.~\ref{Sect5}, we mentioned that the operators
$\ope{M}_{\mu\nu}$ and $\ope{M}_{\mu\nu}^{\text{r}}$ may satisfy the
relations~\eref{5.8} or~\eref{5.9}, respectively. However, in view
of~\eref{5.9} and~\eref{6.2}--\eref{6.4}, the equations~\eref{5.8} may be
valid only when applied to states with vanishing angular momentum (and
momentum --- see~\cite{bp-QFT-momentum-operator}).  Therefore the
relations~\eref{5.8} are, generally, unacceptable. However, in the Lagrangian
formalism, the relations~\eref{5.10} may or may not hold, depending on the
particular Lagrangian employed.


\section {Conclusion}
\label{Conclusion}

	The following results should be mentioned as major ones of this
paper:
	\begin{description}
\item[(i)]
	The generator $\ope{M}_{\mu\nu}^{\text{QM}}$ of (the representation
of) rotations in system's Hilbert space of states is \emph{not} the (total)
angular momentum operator in quantum field theory, but it is closely related
to a kind of such operator (see~\eref{4.12}).

\item[(ii)]
	The rotational angular momentum operator
$\ope{M}_{\mu\nu}^{\text{r}}$ is a generator of (the representation of)
rotations in the space of operators acting on system's Hilbert space of
states. It depends on a second\ndash rank tensor $m_{\mu\nu}$ with constant
(relative to Poincar\'e group) components.

\item[(iii)]
	The canonical total angular momentum operator $\ope{M}_{\mu\nu}$ is,
generally, different from the rotational one. But, if one identifies the
tensor $m_{\mu\nu}$ with the total angular momentum of the system under
consideration, the restrictions of $\ope{M}_{\mu\nu}$ and
$\ope{M}_{\mu\nu}^{\text{r}}$ on the set~\eref{6.5} coincide.

\item[(iv)]
	An operator $\ope{M}_{\mu\nu}^{\text{r}}$, satisfying the commutation
relations~\eref{5.8} of the Lie algebra of the Poincar\'e group, describes a
system with vanishing angular momentum. The operator $\ope{M}_{\mu\nu}$ may or
may not satisfy~\eref{5.10}, depending on the Lagrangian describing the
system explored.

\item[(v)]
	When commutators with field operators are concerned, the operators
$\ope{M}_{\mu\nu}$ and $\ope{M}_{\mu\nu}^{\text{r}}$ are interchangeable
(see~\eref{6.6}).  However, the relations~\eref{4.4} are identities,
while~\eref{5.2} are equations relative to the field operators and their
validity depends on the Lagrangian employed.

\item[(vi)]
	The spin and orbital angular momentum operators (in the Lagrangian
formalism) contain additive terms which are conserved operators and their
sum is the total angular momentum operator. (This result is completely valid
in the classical Lagrangian formalism too, when functions, not operators, are
involved.)

	\end{description}

	As it is noted in~\cite[\S~68]{Bjorken&Drell-2}, the quantum field
theory must be such that the (canonical) angular momentum operator
$\ope{M}_{\mu\nu}$, given in Heisenberg picture via~\eref{2.9}, must satisfy
the Heisenberg relations/equations~\eref{5.2}. This puts some restrictions on
the arbitrariness of the (canonical) energy\ndash momentum tensorial operator
$\ope{\ope{T}}_{\mu\nu}$ and spin angular momentum density operator
$\ope{S}_{\mu\nu}^{\lambda}$ (see~\eref{2.6}, \eref{2.10} and~\eref{2.11}),
obtained, via the (first) Noether theorem, from the system's Lagrangian.
Consequently, this puts some, quite general, restrictions on the possible
Lagrangians describing systems of quantum fields.

	If~\eref{5.2} holds, then, evidently, its r.h.s.\ is a sum of two
parts, the first related to the orbital angular momentum and the second one
--- to the spin angular momentum. This observation suggests the idea to
split the total angular momentum as a sum
	\begin{equation}	\label{c.1}
\ope{M}_{\mu\nu}
=
\ope{M}_{\mu\nu}^{\mathrm{or}}(x) + \ope{M}_{\mu\nu}^{\mathrm{sp}}(x) ,
	\end{equation}
where
$\ope{M}_{\mu\nu}^{\mathrm{or}}(x)$ and
$\ope{M}_{\mu\nu}^{\mathrm{sp}}(x)$
characterize the `pure' orbital and spin, respectively, properties or the
system and
	\begin{align}
			\label{c.2}
[ \varphi_i(x), \ope{M}_{\mu\nu}^{\mathrm{or}}(x) ]_{\_}
& =
\ih \Bigl( x_\mu \frac{\pd}{\pd x^\nu} - x_\nu \frac{\pd}{\pd x^\mu} \Bigr)
\varphi_i(x)
\\			\label{c.3}
[ \varphi_i(x), \ope{M}_{\mu\nu}^{\mathrm{sp}}(x) ]_{\_}
& =
\ih I_{i\mu\nu}^{j} \varphi_j(x)
	\end{align}
from where~\eref{5.2} follows. If we accept the Heisenberg relation for the
momentum operator $\ope{P}_\mu$ of the system,
i.e.~\cite{Bjorken&Drell-2,Bogolyubov&Shirkov,Roman-QFT,Itzykson&Zuber}
	\begin{equation}	\label{c.4}
[ \varphi_i(x), \ope{P}_{\mu} ]_{\_}
=
\ih \frac{\pd}{\pd x^\mu} \varphi_i(x) ,
	\end{equation}
then we can set
	\begin{align}
			\label{c.5}
\ope{M}_{\mu\nu}^{\mathrm{or}}(x)
& =
x_\mu \ope{P}_\nu - x_\nu \ope{P}_\mu
\\			\label{c.6}
\ope{M}_{\mu\nu}^{\mathrm{sp}}(x)
& =
\ope{M}_{\mu\nu} - \bigl( x_\mu \ope{P}_\nu - x_\nu \ope{P}_\mu \bigr) .
	\end{align}
Such a splitting can be justified by the explicit form of $\ope{M}_{\mu\nu}$
for free fields in a kind of 4\ndash dimensional analogue of the
Schr\"odinger picture of motion, which will be considered elsewhere. The
operators~\eref{c.5} and~\eref{c.6}, similarly to~\eref{2.6} and~\eref{2.7},
are not conserved quantities. The physical sense of the operator~\eref{c.5}
is that it represents the angular momentum of the system due to its movement
as a whole. Respectively, the operator~\eref{c.6} describes the system's
angular momentum as a result of its internal movement and/or structure.
Elsewhere we shall present an explicit splitting, like~\eref{c.1}, for free
fields in which the operators $\ope{M}_{\mu\nu}^{\mathrm{or}}(x)$ and
$\ope{M}_{\mu\nu}^{\mathrm{sp}}(x)$ will be conserved ones and will represent
the pure orbital and spin, respectively, angular momentum of the system
considered.


\addcontentsline{toc}{section}{References}
\bibliography{bozhopub,bozhoref}
\bibliographystyle{unsrt}
\addcontentsline{toc}{subsubsection}{This article ends at page}

\end{document}